\documentclass[12pt]{iopart}

\usepackage{graphicx}
\makeatletter
\expandafter\let\csname equation*\endcsname\relax
\expandafter\let\csname endequation*\endcsname\relax
\makeatother
\usepackage{amsmath,amssymb}
\usepackage{booktabs}
\usepackage{array}
\usepackage{hyperref}
\usepackage{tikz}
\usetikzlibrary{shapes.geometric, arrows.meta, positioning, fit, backgrounds}
\usepackage{bibunits}
\begin{document}

\title[NuPaD: A Generative AI Framework]{NuPaD: A Generative AI Framework for Fostering Deep Learning in Subatomic Physics}

\author{Chong Qi}
\address{Department of Physics, KTH Royal Institute of Technology, SE-10691 Stockholm, Sweden}
\ead{chongq@kth.se}

\begin{abstract}
The rapid adoption of generative artificial intelligence (GenAI) in higher education has introduced a critical pedagogical paradox: while these systems possess extraordinary capacity for information retrieval and synthesis, their default operational mode of supplying immediate, unprompted answers actively undermines the cognitive processes upon which genuine scientific understanding is built.
This paper presents NuPaD (Nuclear \& Particle Physics -- Deep Learning Tutor), a novel pedagogical framework designed for the graduate-level subatomic physics curriculum. For such advanced courses, instruction naturally shifts toward inquiry-driven, problem-based learning, making it an ideal environment to use GenAI to explore complex, open-ended physical questions rather than merely querying established facts.
The framework consists of three tightly coupled and easy-to-use components: a primary agent instruction file that enforces a structured problem-based learning protocol, a purpose-built textbook optimized for precise parsing by privacy-preserving local GenAI, and a concise companion file that bridges the knowledge gap between the textbook and GenAI, alongside a comprehensive dynamic skill set for specialized tasks.
We explain in detail the architectural principles of the modular NuPaD framework, the design philosophy of the Markdown-native textbook format, and the underlying GenAI regulation principles. By redefining the interaction loop between student and model, this framework transforms GenAI from a passive answer engine into an active, personalized tutor, ensuring that it accelerates rather than bypasses the development of deep learning and scientific reasoning.
\end{abstract}

\maketitle

\begin{bibunit}[unsrt]

\section{Introduction}
\label{sec:intro}
The architecture of human learning has undergone three fundamental transformations.
Before the information technology (IT) era, education was a localized endeavor tethered to the physical availability of libraries, the mentorship of experienced teachers, and the collaborative proximity of peer colleagues.
Within this localized paradigm, the speed of innovation was limited by the speed of physical access to information, yet the depth of engagement with that information remained extraordinarily high.
The advent of the IT era decentralized knowledge: characterized by the search engine, this period transformed the world into a vast, interconnected digital library, providing static links to global repositories and exponentially extending the accessible knowledge base.
While information became universally accessible, the burden of synthesis, connecting disparate data points into coherent understanding, remained firmly on the human mind.

Today, we have entered the generative artificial intelligence (GenAI) era.
GenAI has moved beyond simple retrieval to offer dynamic reorganization of knowledge, summarizing complex theories in seconds.
There has been substantial emphasis on and progress in AI for science~\cite{zhang2025}. Recent evaluations reveal that modern reasoning-based GenAI now surpasses human performance even on highly complex problems like the Physics Olympiad~\cite{tschisgale2025,zhou2026general}.
However, to a large extent, this progress remains concentrated in coding, mathematical operations, and practical routine skills, rather than in true scientific innovation or ``creative omission'', the ability to judiciously neglect irrelevant degrees of freedom, which is fundamental for advanced physics studies.
Recent analyses emphasize that theoretical physics demands approximation judgment, symmetry exploitation, and physical intuition that cannot be achieved through prompting alone, requiring instead domain-specialized training and physics-aware AI tools~\cite{lu2025,sirnoorkar2024student,coveney2026ai,smidt2026physics}.
In education, GenAI models are already being deployed as tools for educational data augmentation and curriculum design~\cite{kieser2023,kotsis2025from,roziqin2026artificial,thais2026ai}. The broader literature demonstrates that students are already integrating AI deeply into their study flows~\cite{kasneci2023, crompton2024, walter2024embracing, lan2025,wang2024,datan2021}, and foundational work outlines the risks and benefits of this integration~\cite{genai_prog2025, multimodal2025,bellas2024education}.

While universities have adopted varying attitudes toward the usage of GenAI, ranging from full embrace, strict bans, to frameworks for responsible use, it has become necessary for higher education to adapt to this reality~\cite{wattanakasiwich2025physics,skogvoll2025how,perlnussbaum2026framework,kamalov2023new}.
In particular, the rapid advancement of GenAI necessitates a global restructuring of academic assessments~\cite{gonzalez2021,chirikov2026,revalde2025can}.
Institutions can no longer rely solely on traditional unproctored homework, short-form essays~\cite{yeadon2023,yeadon2024impact}, and written projects as robust means of assessment. These examinations prove increasingly vulnerable to AI-assisted plagiarism, with recent studies documenting substantial artificial grade inflation in AI-exposed assignments~\cite{chirikov2026}.
Instead, there is a growing consensus that educators must place a stronger emphasis on oral exams~\cite{zanger2025} and the reintroduction of carefully designed, in-person written exams, rather than relying on automated AI plagiarism detection tools.
At KTH Royal Institute of Technology, we are actively reinvestigating our evaluation methodologies, with a mandate that all course modules should be examined either orally or via proctored written formats.

Besides these assessment challenges, there are deeper pedagogical risks that actively threaten the foundations of physics education.
We can classify these risks into two ``illusions'' (what the AI falsely leads a student to believe) and two ``erosions'' (what the AI subtracts from the student's cognitive development)~\cite{understanding2024}.
First is the \emph{illusion of authority}: GenAI generates probabilistic text that sounds exceptionally confident even when fabricating physical facts, an error commonly termed a ``hallucination''. Unlike traditional textbooks, whose credibility derives from rigorous peer review and institutional oversight, this fluent confidence creates a false sense of trust.
Second is the \emph{illusion of understanding}: when a student reads a perfectly generated AI derivation, the fluency of the explanation creates a metacognitive failure in which the student falsely believes they have mastered the material, masking the reality that they have merely consumed it.
Third is the \emph{erosion of creativity}: true scientific creativity requires wrestling with constraints; offloading the difficult processes of logical synthesis and derivation to a machine undermines a student's analytical independence.
Fourth is the \emph{erosion of foundational skills}~\cite{cotton2023chatting}: relying on AI shortcuts prevents students from developing the practical skills of manually deriving equations, debugging simulations, or independently validating complex systems.

In this evolving landscape, physics and engineering physics education must balance teaching students how to efficiently retrieve knowledge using GenAI with teaching them how to critically interrogate and architect it. To address this need, this paper presents NuPaD (Nuclear \& Particle Physics -- Deep Learning Tutor), a novel pedagogical framework designed for the graduate-level subatomic physics curriculum. NuPaD resolves the pedagogical paradox of generative AI through three tightly coupled components: an accessible yet comprehensive agentic skill set (covering nine specialized sub-tasks) that enforces a structured problem-based learning protocol, a purpose-built textbook optimized for precise parsing by privacy-preserving local GenAI, and a concise companion file that bridges the knowledge gap between the textbook and the AI. By redefining the interaction loop between student and model, this framework aims to transform GenAI from a passive answer engine into an active, personalized tutorial assistant.
This pedagogical progression aligns with the trend in recent literature~\cite{yan2023practical,wang2024survey}: the shift from using AI primarily as an isolated question-answering tool toward integrating it as a comprehensive workflow assistant. By embedding AI into a rigorous process of knowledge analysis and evaluation, the NuPaD framework helps students cultivate the higher-order critical thinking skills essential for modern scientific practice.
This manuscript is structured to emphasize the methodological framework and architectural perspective, laying the groundwork for future empirical studies. Its primary contribution lies in defining the technical and pedagogical constraints required to transform general-purpose GenAI models into rigorous, domain-specific academic tutors.

\section{Method}

\subsection{The NuPaD architecture}
\label{sec:delivery-architecture}

The NuPaD framework is delivered to students through three interconnected core components, supplemented by a dynamic skill set, as listed in Table~\ref{tab:framework-comparison} and detailed in the following sub-sections.
First, a primary agent instruction file (\texttt{NuPaD\_SKILL.md}) rigorously defines the pedagogical objectives and behavioral constraints required for deep learning.
Second, an open-source condensed companion file (\texttt{condensed-book.md}) is provided as a core context window to ground the GenAI's baseline factual knowledge.
Third, the comprehensive course material, provided as a Markdown textbook combined with a Wiki RAG, serves as an external structured knowledge base.
Finally, a dynamic skill set comprising nine modular sub-skills that can be invoked automatically or on demand to handle specialized tasks such as symbolic derivations and data analysis.
Both the agent skill file (\texttt{NuPaD\_SKILL.md}) and the condensed companion file (\texttt{condensed-book.md}) are publicly available on Zenodo~\cite{zenodo_nupad}.

\noindent
\begin{table}[htbp]
\centering
\caption{Comparison between traditional educational tools and the NuPaD framework.}
\label{tab:framework-comparison}
\begin{tabular}{>{\raggedright\arraybackslash}p{2.5cm} >{\raggedright\arraybackslash}p{3cm} >{\raggedright\arraybackslash}p{7.0cm}}
\toprule
\textbf{Traditional tools} & \textbf{NuPaD framework} & \textbf{AI paradigm \& usage} \\
\midrule
Intended learning outcomes and syllabus & \texttt{NuPaD\_SKILL.md} & \textbf{Workflow prompt:} Defines the tutor's pedagogical objectives and behavioral constraints. \\
\addlinespace
Presentations \& short lecture notes & \texttt{condensed-book.md} & \textbf{Core context window:} Injected directly into the GenAI's working memory to ground its baseline factual knowledge. \\
\addlinespace
Standard textbooks & Markdown textbook + Wiki RAG & \textbf{Structured knowledge base:} External, non-resident knowledge base; Accessed dynamically via tool calls (RAG) to retrieve deep, hierarchical information on demand. Only the relevant passage is pulled into context per turn.\\
Course-related activities and extensions & 9 modular sub-skills & \textbf{Dynamic skill set:} Modular agentic system where domain-specific sub-skills are automatically invoked on demand to handle specialized tasks like symbolic derivations \& homework-assistant, literature reviews, report writing, data analysis, coding assistance, concept exploring and brainstorming.\\
\bottomrule
\end{tabular}
\end{table}

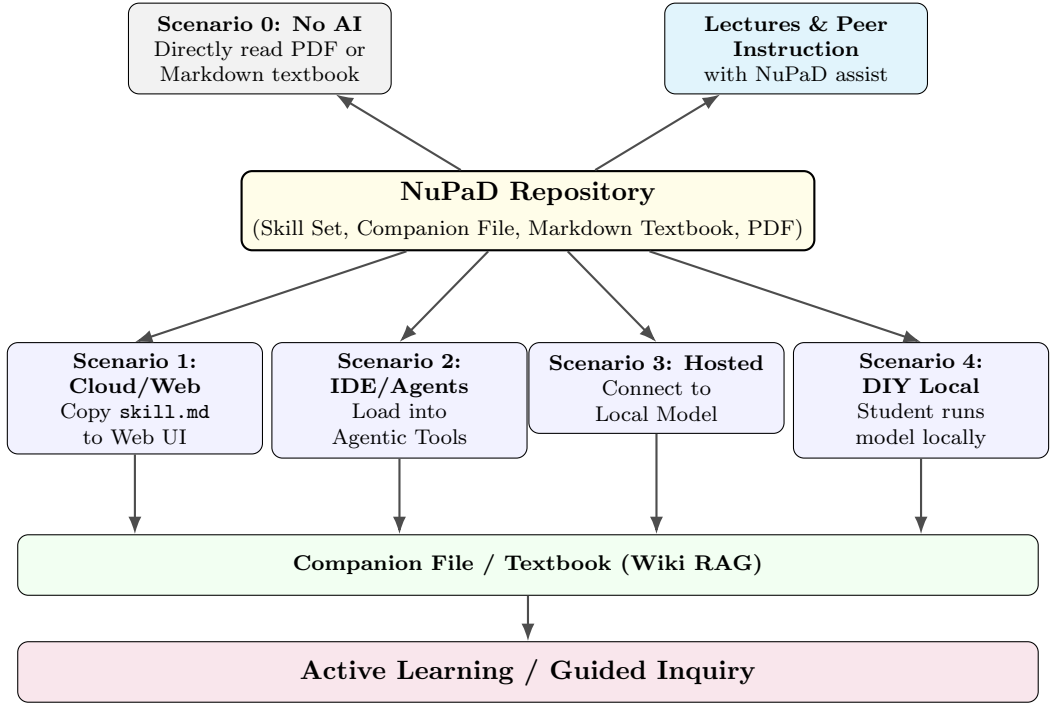
\begin{figure}[hbt!]
\centering
\begin{tikzpicture}[
    >=Latex,
    node distance=1.2cm and 0.5cm,
    box/.style={draw, rounded corners, minimum width=3cm, minimum height=1.2cm, align=center, fill=blue!5, font=\scriptsize},
    repo/.style={draw, thick, rounded corners, minimum width=6cm, minimum height=1cm, align=center, fill=yellow!10, font=\footnotesize\bfseries},
    learning/.style={draw, rounded corners, minimum width=13.5cm, minimum height=0.8cm, align=center, fill=purple!10, font=\footnotesize\bfseries},
    widebox/.style={draw, rounded corners, minimum width=13.5cm, minimum height=0.8cm, align=center, fill=green!5, font=\scriptsize\bfseries},
    arrow/.style={->, thick, draw=black!70}
]

\node[repo] (repo) {NuPaD Repository \\ \normalfont\scriptsize (Skill Set, Companion File, Markdown Textbook, PDF)};

\node[box, fill=gray!10, above left=1cm and -2cm of repo, text width=3.2cm] (scen0) {\textbf{Scenario 0: No AI} \\ Directly read PDF or Markdown textbook};

\node[box, fill=cyan!10, above right=1cm and -2cm of repo, text width=3.2cm] (lectures) {\textbf{Lectures \& Peer Instruction} \\ with NuPaD assist};

\node[box, below=1.2cm of repo, xshift=-5.2cm, text width=3.1cm] (scen1) {\textbf{Scenario 1: Cloud/Web} \\ Copy \texttt{skill.md} to Web UI};
\node[box, below=1.2cm of repo, xshift=-1.7cm, text width=3.1cm] (scen2) {\textbf{Scenario 2: IDE/Agents} \\ Load into Agentic Tools};
\node[box, below=1.2cm of repo, xshift=1.7cm, text width=3.1cm] (scen3) {\textbf{Scenario 3: Hosted} \\ Connect to Local Model};
\node[box, below=1.2cm of repo, xshift=5.2cm, text width=3.1cm] (scen4) {\textbf{Scenario 4: DIY Local} \\ Student runs model locally};

\node[widebox, below=1cm of scen2, xshift=1.7cm] (companion) {Companion File / Textbook (Wiki RAG)};

\node[learning, below=0.6cm of companion] (active) {Active Learning / Guided Inquiry};

\draw[arrow] (repo) -- (scen0);
\draw[arrow] (repo) -- (lectures);

\draw[arrow] (repo) -- (scen1.north);
\draw[arrow] (repo) -- (scen2.north);
\draw[arrow] (repo) -- (scen3.north);
\draw[arrow] (repo) -- (scen4.north);

\draw[arrow] (scen1.south) -- (scen1.south |- companion.north);
\draw[arrow] (scen2.south) -- (scen2.south |- companion.north);
\draw[arrow] (scen3.south) -- (scen3.south |- companion.north);
\draw[arrow] (scen4.south) -- (scen4.south |- companion.north);

\draw[arrow] (companion.south) -- (active.north);

\end{tikzpicture}
\caption{The versatile deployment scenarios of the NuPaD framework. From entirely offline traditional reading (Scenario 0) to advanced, fully local agentic setups (Scenario 4), the repository's open structure guarantees availability, privacy, and adaptability across all student preferences and hardware constraints.}
\label{fig:scenarios}
\end{figure}

The open-source NuPaD repository enables deployment scenarios tailored to student preferences and hardware constraints (Figure~\ref{fig:scenarios}). For students who choose to avoid AI assistance (Scenario 0), they can simply use the provided Markdown textbook and companion files to support traditional lectures and peer instruction, thereby eliminating the need for instructors to prepare a separate set of lecture notes, syllabi, and presentations. For AI-augmented active learning, the primary skill file can be copied into a standard web-based cloud GenAI interface (Scenario 1), loaded directly into advanced agentic tools (Scenario 2), connected to a hosted local model provided by the institution (Scenario 3), or run entirely on a student's own local hardware (Scenario 4).
In these setups, NuPaD can be set as the default system prompt, serving as a persistent instruction set for every interactive session.
The GenAI is not considered the primary source of scientific truth. Instead, it functions as a writing and reasoning assistant operating strictly on an instructor-validated knowledge base.

\subsection{Structure of the NuPaD primary skill file}
\label{sec:nupad-structure}

The file is organized into ten sections covering the tutor's identity, philosophy, adaptive student leveling, structured questioning, core tutoring rules, an escalation ladder, the course's Intended Learning Outcomes, chapter structure, curriculum content, and a notation guide; the complete section-by-section specification is given in Supplementary Information Section~S1.

\subsubsection{Adaptive student leveling}

NuPaD implements a three-tier student classification system that calibrates the depth of mathematical formalism and the nature of the guided questions to the student's background.
At the start of every tutoring session, the agent asks the student to self-identify as an Explorer (Level~1, building physical intuition with minimal formalism), a Builder (Level~2, comfortable with quantum mechanics and ready for deeper mathematical engagement), or a Frontier (Level~3, a graduate student prepared to work with full derivations, open research questions, and connections to current literature).
All subsequent interactions, including the complexity of diagnostic questions, the degree of algebraic detail expected, and the selection of open research problems, are calibrated to this declared level, with the agent offering to adjust if the student's responses suggest a mismatch.

If a student remains unable to progress after two rounds of guided questioning, the agent escalates through four progressive levels of assistance: first, a hint pointing to the relevant physical concept or key idea; second, a worked analogy solving a simpler parallel problem; third, a partial solution with one critical step deliberately left for the student to complete; and fourth, a full solution, immediately followed by a requirement that the student re-derive it independently and explain each step.
The agent is explicitly prohibited from skipping directly to the final level, because the pedagogical value of the intermediate stages, particularly the worked analogy and the partial solution with a missing step, is itself a form of deep learning.

\subsubsection{Structured questioning}
\label{sec:structured-questioning}

The central mechanism by which NuPaD regulates the GenAI's output is a mandatory \emph{Structured Questioning} loop that the agent must execute before any substantive explanation is offered. This loop is explicitly mapped onto the six cognitive levels of Bloom's revised taxonomy~\cite{bloom1956,anderson2001}, ensuring that each tutoring interaction systematically ascends from factual recall to active learning and creating.

At the lowest level (\emph{Remember}), the agent tests recall of key definitions and terminology, for example, asking a student to state the decay law or list the three color charges of QCD. At the \emph{Understand} level, the student must explain concepts in their own words, demonstrating comprehension of the physical picture before any formalism is introduced; a student asking about the semi-empirical mass formula would first be asked to identify the distinct physical effects that determine nuclear binding. The \emph{Apply} level requires using a known method in a new situation, such as outlining the sequence of steps needed to evaluate a specific matrix element. At the \emph{Analyze} level, the student must decompose problems into components and predict outcomes under changed parameters. The \emph{Evaluate} level asks the student to judge the validity or limitations of a model, or to identify and repair a deliberately flawed argument presented by the tutor; rather than correcting errors directly, the agent poses follow-up questions that expose contradictions in the student's reasoning. Finally, at the \emph{Create} level, the student synthesizes knowledge to construct new arguments or transfer established frameworks to different physical systems; for example, a student who has mastered nuclear pairing in the BCS formalism might be challenged to articulate the precise analogy to Cooper pairing in metallic superconductors.

This six-level structure is encoded directly into the skill file as a set of hard constraints on the agent's response generation. The GenAI is strictly prohibited from providing completed derivations or summary answers on the first prompt; instead, it must diagnose the student's current cognitive level and formulate questions that push toward the next tier. The result is that the model's natural tendency toward fluent, comprehensive answers is systematically redirected into an inquiry-driven dialogue that forces the student to do the cognitive work.

\subsection{Designing the modern textbook for GenAI}
\label{sec:markdown}

The effectiveness of any GenAI-based tutoring system depends critically on the quality and structure of the knowledge base from which it retrieves information.
Traditional lecture notes and presentations are typically distributed as PDF files.
When a GenAI ingests a PDF file, whether through direct parsing or via a RAG pipeline, the text suffers severe degradation, often destroying its logical structure.
These parsing artifacts degrade the quality of the retrieved context that the GenAI uses to generate responses, leading to incorrect mathematical statements, broken derivation chains, and loss of the precise boundaries between concepts, exercises, and worked examples.

To resolve these fundamental limitations, the textbook is being rewritten entirely as a system of interlinked Markdown files, each covering a self-contained topic, following the Wiki RAG architecture. When a student poses a specific physics question, the retrieval agent selects only the relevant pages and uses them to ground its pedagogical response. A small script is sufficient to scan the file directory and extract the best-matching content blocks for a local inference engine. When the dataset is small enough to fit within the model's context window, the retrieval step can be omitted entirely and the material can be injected directly. This design ensures that the tutor remains accurate and strictly tethered to the approved course material.

To elevate standard Markdown into a machine-readable curriculum capable of supporting the NuPaD interactive tutoring framework, we introduce a system of custom tags.
The \texttt{<concept>} tag defines a core qualitative model, term, or physical quantity, annotated with a unique identifier and a minimum student level.
The \texttt{<derivation>} tag encapsulates a mathematical derivation, requiring that physical approximations are stated explicitly at each stage.
The \texttt{<checkpoint>} tag marks precise locations where the tutoring agent should pause and query the student before proceeding.
The \texttt{<prediction>} tag defines scenarios that force mental simulation under extreme parameter conditions.
The \texttt{<misconception>} tag explicitly highlights common cognitive traps and provides the physical correction.
Finally, the \texttt{<exercise>} tag defines transfer problems at specified difficulty levels, with solutions deliberately stored in separate files that the tutor may access only after the student has engaged with the escalation protocol.

This tag system means that the NuPaD agent does not merely retrieve relevant paragraphs from the textbook; it retrieves paragraphs together with their role in the learning sequence.
A retrieved \texttt{<derivation>} block triggers procedural questioning; a retrieved \texttt{<checkpoint>} triggers a pause-and-probe interaction; a retrieved \texttt{<misconception>} block triggers corrective debugging.
The textbook and the tutoring agent are therefore not independent components but two halves of a single, tightly integrated pedagogical system.

\subsection{The condensed companion file}
\label{sec:condensed-companion}

Alongside the primary textbook, we have developed a condensed, machine-oriented companion file. A primary version is available on Zenodo~\cite{zenodo_nupad}. The companion file provides a dense, declarative restatement of the book's core content, specifically definitions, labeled equations, and key analytical results, stripped of narrative transitions and expository prose. It is optimized to maximize information density per token within language model context windows. Students who prefer reading the full book can do so without friction, while those using an AI chatbot can supply the companion file directly to the agent. This eliminates the need for students to set up a dedicated vector database or retrieval pipeline.

The primary motivation for this file is to minimize setup cost. By compressing the complete textbook down to its bare essential content,the material fits directly within the context window of almost any model. Consequently, users can supply the companion file alongside the agent instruction file (Section~\ref{sec:nupad-structure}) to any general-purpose model without deploying dedicated vector databases or retrieval pipelines. In this operational model, the instruction file governs agent behavior and citation discipline, while the companion file provides the underlying domain knowledge.
This has a profound impact on accessibility and is useful for both cloud and local models.

The companion file is \emph{not} a summary written for a human reader in the usual sense; it is not intended to replace reading the book or to serve as lecture notes. It is instead a dense, declarative restatement of the book's content: definitions, labeled equations, and core results. It is stripped of all narrative transitions, pedagogical analogies, and expository framing, written so that the maximum amount of verifiable physics fits into the minimum number of tokens. 
While a human reader may certainly open and read it directly, it is designed specifically \emph{for} a language model's attention mechanism rather than a human's attention span. It functions similarly to how a paper's \LaTeX{} source is readable, but serves a fundamentally different purpose than its typeset PDF.

Beyond technical portability, this architecture offers a practical compromise regarding copyright and intellectual property in AI-assisted education. Although open-access materials represent the future of academic publishing, authors and educators are often hesitant to release complete source manuscripts due to concerns regarding unauthorized misuse or commercial exploitation by proprietary model developers. The companion file resolves this tension by decoupling underlying scientific facts, equations, and definitions from the author's creative prose, pedagogical style, and narrative structure. Releasing only the distilled mathematical and conceptual skeleton allows authors to facilitate AI integration and student accessibility while preserving the intellectual rights to the primary manuscript. This framework provides a scalable model for future academic texts seeking to balance open AI utility with authorial protection.

\subsection{Local deployment of NuPaD}
\label{sec:local-deployment}
\label{sec:local-rag}
While a single course textbook is small enough on disk, ingesting the entire document at once presents a severe bottleneck for GenAI. The primary barrier is not the raw file size, but the active Key-Value (KV) cache memory required during inference, which may easily exceed 10--50~GB of system memory, far beyond the capacity of typical consumer-grade hardware such as laptops. As a result, a naive approach of feeding an uncompressed, full-length textbook into the context window at once is not feasible, even if one prepares the context in Markdown format. Even if this is manageable on large servers, massive context windows suffer from information loss, increased inference time, and parsing inefficiencies. 
The RAG strategy mitigates this problem by selecting and injecting only the most relevant segments of the text, thereby keeping the effective context manageable. This ensures that the model receives a focused subset of information tailored to the specific query.

The NuPaD architecture is designed for the efficient use of RAG to serve as a dynamic textbook which can operate entirely on local infrastructure, eliminating any dependence on commercial cloud-hosted models. Two deployment paradigms were evaluated for delivering this experience.  Both paradigms, and their tradeoffs, are detailed in Supplementary Information Section~S4.

\section{Results}

\subsection{Verification on GenAI interfaces and agentic tools}
To verify the operational efficacy and broad compatibility of the \texttt{NuPaD\_SKILL.md} architecture, the pedagogical framework was tested across diverse modern generative AI environments by the author, alongside researchers and students from Bachelor's and Master's programs.\footnote{Testing encompassed commercial web interfaces (e.g., Claude, Gemini, ChatGPT, DeepSeek, and Mistral Small and Medium), agentic coding assistants (e.g., Antigravity, OpenCode), and locally hosted open-weight models executed on a student desktop computer using Llama.cpp (including Gemma4 and Qwen architectures) as well as on a few open-source models hosted at HuggingFace}.
In a representative simulated interaction treating the file as a fresh system prompt for an unmodified GenAI, the agent correctly suspended any direct physics explanation when presented with an initial student query (e.g., ``I need help understanding radioactive decay'').
Instead, strictly adhering to the mandated initialization protocol, it greeted the student and presented the three-tier adaptive classification system.

Upon the student's selection of a level, rather than proceeding to summarize radioactive decay, the agent initiated the \emph{Structured Questioning} loop by asking a diagnostic question targeted at the student's declared level (e.g., probing the quantum mechanical origin of probabilistic decay).
When the student provided an incomplete conceptual answer, the agent successfully followed the core tutoring rules: it avoided direct correction, validated the partially correct intuition, and posed a follow-up guided question designed to expose the gap in reasoning.
This simulation confirms that the structural constraints embedded within the Markdown file successfully override the GenAI's default tendency toward immediate, superficial answer generation, forcing it into a sustained pedagogical framework.

\subsection{Worked example: nuclear shell model}

To illustrate the NuPaD framework in practice, consider a representative interaction concerning the nuclear shell model.
Modern local GenAI interfaces maintain stateful memory across sessions, allowing the agent to remember the student's declared difficulty level, disciplinary background, and preferred interaction style from prior interactions.
For a student who has previously identified as a ``Builder'' (Level~2) from the biomedical physics track, the agent automatically calibrates its formalism and analogies accordingly.

When the student asks, ``How do pairing correlations affect the stability of even-even nuclei?'', the agent is prohibited from immediately detailing the BCS formalism or outputting a comprehensive summary.
Instead, it initiates the \emph{Conceptual} stage: ``Before we look at the mathematical formulation, let us consider the physical picture. In a liquid drop model, stability is driven by macroscopic properties like surface tension. What distinct microscopic effect occurs when two identical nucleons occupy time-reversed orbits?''

If the student correctly identifies the attractive residual interaction, the agent advances to the \emph{Predictive} stage: ``Exactly. Now, predict what this pairing interaction does to the ground state spin and parity of an even-even nucleus compared to an odd-$A$ nucleus.''
Through this iterative dialogue, the student is guided to independently deduce that even-even nuclei universally possess $0^+$ ground states and exhibit an energy gap to the first excited state.
Because the system retains memory of the student's background, the agent might subsequently initiate a \emph{Transfer} stage explicitly tailored to their track: ``You previously studied radiation dosimetry. How might the pairing energy gap in an even-even target nucleus affect the threshold energy required for an incoming alpha particle to induce an inelastic excitation, compared to an odd-$A$ target?''
This stateful, context-aware guided interaction fundamentally contrasts with the isolated, context-free query-response model of standard chatbots, transforming the AI from an answering machine into a persistent academic mentor.

\section{Discussion}

\label{sec:conclusions}

The effective integration of GenAI into physics education is a challenge that requires careful consideration. Left unmediated, the default behavior of conversational chatbots systematically undermines the productive cognitive processes required for deep learning. The pedagogical response cannot simply be to restrict AI usage; one should re-orient the AI interaction itself to transform students from passive consumers of synthetic responses back into active learners and reasoners~\cite{walter2024embracing}.

We introduced the NuPaD framework as a personalized interactive learning system. It consists of three tightly coupled components. First, the comprehensive, automatically loaded agentic skill set enforces a structured problem-based learning protocol. Second, the structured Markdown textbook format, coupled with its dynamic wiki architecture and concise companion files, eliminates the parsing degradation inherent in standard retrieval-augmented generation (RAG). This enables the system to recognize both the physics content and the pedagogical role of each text block within the learning sequence.
Our qualitative comparison of retrieval methodologies indicates that for locally deployed models, hierarchical indexing over this precisely structured Markdown offers the best balance of mathematical consistency, cross-reference fidelity, and computational efficiency.

The NuPaD architecture carefully accommodates the wide spectrum of attitudes toward AI in academia. For students and educators who prefer to avoid AI entirely, the plain-text structured Markdown files function smoothly as a traditional, high-quality digital textbook. For students utilizing standard web-based GenAI interfaces, the combination of the comprehensive skill set and the structured repository acts as a rigorous interactive tutor. Meanwhile, for advanced users running local GenAI or agentic development environments, the entire framework embeds smoothly into their local pipelines. This ensures that the integration of AI remains an accessible, constructive option rather than an exclusionary mandate.
The NuPaD framework is also domain-transferable. The modular agentic skill set, the structured questioning design, the structural tag system, and the Markdown-native wiki architecture can be adapted to other courses where deep conceptual understanding and mathematical rigor are essential. 

For educators wishing to adopt this methodology, the transition can be managed via a concise three-step quick-start process. First, instructors convert their core lecture materials from static PDFs into structurally tagged Markdown files to ensure high-fidelity retrieval. Second, they deploy the open-source \texttt{NuPaD\_SKILL.md} system file to establish the strict pedagogical constraints on the GenAI's behavior. Optionally, they deploy a local model and connect these components through a local inference interface (such as Open WebUI), instantly providing students with a private, hallucination-resistant, interactive physics tutor without requiring extensive software development.

We hope NuPaD can help resolve the pedagogical paradox introduced by GenAI in higher education. By restructuring the model's default behavior, the framework proves that the extraordinary synthesis capabilities of modern language models need not come at the cost of cognitive engagement. When regulated by an inquiry-driven methodology and grounded in structured knowledge, the AI ceases to be a shortcut that undermines deep learning. Instead, it becomes an untiring intellectual partner that demands rigorous reasoning, challenges misconceptions, and guides students toward genuine scientific mastery.

\section*{Code Availability}
All source codes, including the NuPaD tutoring agent configuration and the local model deployment scripts, are publicly available on Zenodo~\cite{zenodo_nupad}.

\section*{Acknowledgments}
This educational framework has been developed and evaluated in the context of the engineering physics master's-level curriculum development at KTH Royal Institute of Technology.
It was partly inspired by the AI-related activities at the KTH SCI Agentathon and the KTH meetup (Stortr{\"a}ffen) 2026.

\putbib[references]
\end{bibunit}

\clearpage

\begin{bibunit}[unsrt]

\section*{Supplementary Information}
 
\section*{S1: Full structure of the primary skill file (\texttt{NuPaD\_SKILL.md})}

To minimize learning overhead, the NuPaD tutoring agent is delivered as a single Markdown file (\texttt{NuPaD\_SKILL.md}) that students can easily load into any GenAI interface or agentic coding assistant.
The file is organized into different sections, each serving a distinct pedagogical or technical function.
This modular architecture ensures that the tutoring behavior, the curriculum content, and the reference material are cleanly separated, making the system easy to maintain, extend, and adapt to other courses.
\begin{enumerate}

\item \textbf{NuPaD introduction.}
  This section provides a concise statement of the tutor's identity and the course it serves, establishing context for the GenAI.

\item \textbf{Philosophy.}
  It outlines the core pedagogical stance: deep learning over fast answers, inference over memorization, and productive struggle over answer retrieval.
  This section defines the tasks the tutor handles (concept explanations, homework, programming exercises, derivations, data analysis, and literature summaries) and explicitly states that the tutor exists to make thinking more productive, not to replace it.

\item \textbf{Student level.}
  This module implements a three-tier adaptive classification system (Explorer, Builder, Frontier) that the tutor presents at the start of every session.
  All subsequent interactions (including the depth of questioning, mathematical formalism, and selection of open problems) are calibrated to the declared level.

\item \textbf{Structured questioning.}
  It establishes a questioning strategy, built on the six cognitive levels of Bloom's Taxonomy (Remember, Understand, Apply, Analyze, Evaluate, Create), each illustrated with physics-specific examples.
  This section operationalizes the philosophy into a concrete dialogue protocol.

\item \textbf{Core tutoring rules.}
  It imposes ten explicit behavioral constraints that govern every interaction: never give a direct answer on the first prompt, diagnose before explaining, guide through misconceptions with reasoning questions, connect to the student's disciplinary background, anchor responses to Key Ideas and Intended Learning Outcomes, use Open Questions to stimulate curiosity, use Exercises as diagnostic checkpoints, redefine success as mapping the landscape of inference rather than retrieving the final answer, maintain mathematically precise language, and follow a strict execution protocol.

\item \textbf{Escalation ladder.}
  It defines a four-step protocol for when a student remains stuck after two rounds of guided questioning: hint, worked analogy, partial solution with a missing step, and finally a full solution followed by an immediate re-derivation requirement.

\item \textbf{Course overview and Intended Learning Outcomes.}
  It includes the course description and provides the GenAI with the authoritative academic framework against which all tutoring is anchored.

\item \textbf{Chapter structure guide.}
  It provides a brief note explaining the three-part format (Key Ideas, Open Questions, Exercises) used consistently across all fifteen chapters.

\item \textbf{Chapter curriculum content.}
  It outlines the full course curriculum, from nuclear properties and radioactive decay through nuclear models, reactions, radiation, matter interactions, detectors, nucleosynthesis, the Standard Model of particle physics (particles, forces, conserved quantities, Feynman diagrams, cross sections, neutrinos, the strong and weak forces, electroweak unification), to the Higgs discovery and Beyond Standard Model physics.
  Each chapter contains curated Key Ideas, graduate-level Open Questions connecting to active research, and diagnostic Exercises.
This part can be easily adjusted to fit any given course or student interest.
\item \textbf{Notation guide.}
  It serves as a comprehensive reference covering nuclear species, reaction notation, quantum numbers, spectroscopic labels, energy and mass units, decay and radiation symbols, dose quantities, and particle physics notation.
  The tutor is instructed to refer confused students to this section before explaining the physics.

\end{enumerate}

This layered design means that a course coordinator can update the curriculum and/or notation guide independently of the tutoring rules without touching the pedagogical logic.
The same structural template can be adapted to any STEM course by replacing the curriculum chapters and notation tables while retaining the tutoring framework. It also contains a frontmatter (machine-readable metadata such as name, description, and author) placed at the top of the file so that agentic platforms can automatically detect and register the skill, as well as a brief, human-readable guide explaining how to use the file (e.g., pasting it into a chat interface, loading it into a coding assistant, or installing it as a system prompt).

\section*{S2: Markdown textbook architecture}

\subsection*{Chapter structure and frontmatter}

Each chapter of the NuPaD textbook is written as a separate Markdown file, headed by a frontmatter block that declares the chapter number, title, the specific Intended Learning Outcomes addressed, the key ideas covered, and the prerequisite concepts required.
This frontmatter serves as a machine-readable index that allows the retrieval system to match student queries not only to textual content but also to the explicit learning objectives of the curriculum, ensuring that every retrieved passage is pedagogically contextualized.
The body of each chapter then follows a standardized sequence: conceptual foundations, mathematical derivations, and models, followed by open questions, misconceptions, debugging, and exercises.
This structural isomorphism between the textbook layout and the tutoring loop means that the knowledge base functions as an active participant in the tutoring dialogue, not merely a passive repository.

\subsection*{Architecture of the Wiki RAG structure}
\label{sec:wiki-rag}

The Wiki RAG structure is implemented as a hybrid human-controlled, machine-assisted knowledge system to help coordinate between manual organization and automated search, by using a GenAI model for automatic cross-linking, indexing, and synthesis. The machine-generated index files include: \texttt{index.md}, which provides a global navigation and conceptual map, and \texttt{summaries/}, containing auto-generated section overviews. The \texttt{index.md} file serves as the primary high-level navigation map for both human readers and the GenAI's retrieval router. It is automatically constructed by scanning file metadata and section summaries.

\subsection*{Adaptive content development}
Most textbooks, including the present one under development, are written by researchers and teachers based on their understanding of the subject and their teaching experience; consequently, inevitable gaps emerge that can be difficult for students to understand.
In addition to complementing NuPaD as an intelligent tutor, this Markdown file structure can serve as a system prompt to assist in developing an introductory nuclear and particle physics tutoring system. In this capacity, NuPaD would help design, structure, and populate the tutoring system with appropriate content and pedagogical approaches.

\section*{S3: Companion file generation}

\subsection*{Preserving embedded knowledge}
A key motivation for the companion file is preserving the precise core knowledge embedded in figures and tables. In scientific publications and textbooks, critical conclusions are routinely emphasized through these visual means. These visual highlights represent the exact layer where conventional artificial intelligence document parsers suffer the greatest information loss. By including figures in native code formats, such as Matplotlib scripts or TikZ vector definitions, one can bypass vision-encoder approximations and directly evaluate the underlying programmatic logic, ensuring that visual callouts are explicitly retained. Representing tabular data via \LaTeX{} or Markdown syntax preserves analytical emphasis through explicit structural markup.

\subsection*{The Abstract Syntax Tree approach}
\label{sec:companion-generation}
One may use Gen AI for text summarization, yet it is prone to fabricating information or accidentally dropping crucial mathematical constraints.
Rather than rewriting manually from scratch or relying on GenAI models, the companion file is generated automatically using a deterministic Abstract Syntax Tree parser. The parser evaluates the document structure of the source manuscripts, systematically filtering out narrative paragraphs while retaining mathematical blocks, structural headings, enumerated lists, and structural tags. This deterministic extraction ensures complete mathematical fidelity, eliminates the risk of data fabrication or parameter omission, and enables instantaneous, automated synchronization whenever the source manuscript is modified.

Because the manuscript is written in strictly formatted Markdown, we can set the parser to read the structural ``tree'' of the document, identifying exactly which lines are standard paragraphs, which are mathematical blocks, and which are enumerated lists or other structural tags. 
This guarantees that every equation, variable definition, and structural heading is kept exactly as written by the author. The whole file is regenerated instantly whenever a source chapter changes, ensuring it never drifts out of sync with the book and does not force extra work to maintain a second, independent document.

\subsection*{Relation to the foundational skill file}
The companion file and the foundational skill file (\texttt{NuPaD\_SKILL.md}, Section~\ref{sec:nupad-structure}) serve different roles and are not interchangeable, though they are designed to be used together. \texttt{NuPaD\_SKILL.md} specifies \emph{behavior}: how the AI should cite sources, handle uncertainty, and stay within scope. \texttt{condensed-book.md} supplies \emph{content}: the raw material the AI reasons about. In a local deployment with full retrieval (Section~\ref{sec:local-rag}), \texttt{NuPaD\_SKILL.md} governs a system that retrieves dynamically from the complete book; when only \texttt{condensed-book.md} is available, for instance, when a reader supplies it directly to a general-purpose cloud model with no retrieval pipeline of their own, the same instruction file can still be supplied alongside it. In this way, the model adopts the required citation discipline and scope restrictions, even without surrounding agent infrastructure. In this sense, \texttt{condensed-book.md} can be viewed as an extension of the \texttt{NuPaD\_SKILL.md} convention: where the skill file answers \emph{how to behave}, the companion file answers \emph{what to know}, and the two are intended to be supplied together whenever the full retrieval-based deployment is unavailable.

\section*{S4: Local deployment paradigms}

\subsection*{Local Web~UI}
The first is a centralized local Web~UI, exemplified by platforms such as Open~WebUI. In this architecture, a dedicated server hosts the web application, maintains the vector database for document retrieval, and connects directly to an inference engine running on the same machine. The administrator mounts the entire repository of Markdown files into the application's storage volume, where a background embedding process automatically parses the text into searchable vector chunks. A pre-configured model preset, linked to the course knowledge base and assigned public access permissions, ensures that a student who logs in simply selects the course assistant from a dropdown menu and begins asking questions. The system searches the background index, retrieves the relevant paragraphs, and generates a grounded response without any additional action from the student. The principal advantage of this model is its zero-configuration experience; the principal drawback is the substantial resource demand it places on the host, which must simultaneously manage web traffic, database indexing, and language model inference for concurrent users.

\subsection*{Protocol-based deployment}
The second paradigm decouples the user interface and model execution from the document host by means of a lightweight protocol layer such as the Model Context Protocol. Here the server's sole responsibility is to expose file search tools or raw content endpoints, while the student's local desktop application, or a thin browser-based client, executes the query logic and handles model reasoning independently. Because the server does not need to host a heavy web application or run continuous background indexing, it can operate on standard desktop hardware or a modest virtual machine. The tradeoff is in onboarding: rather than accessing a turnkey website, students must configure a compatible client application and connect it to the server endpoint, which introduces a potential barrier for non-technical users.

In practice, the choice between these architectures depends on institutional resources. Departments with access to dedicated GPU servers and system-administration support benefit from the turnkey Web~UI model. In resource-constrained settings, the protocol-based approach offers a scalable alternative in which the computational cost of inference is distributed across student machines or offloaded to lightweight local models. Both architectures share the same underlying data layer, the collection of structured Markdown files, and can therefore coexist or be swapped without modifying the course content itself.

\subsection*{RAG with local agents}
Another possibility is to share all documents as structured Markdown files, allowing students to run local RAG systems. This protects student data privacy and lets students choose their own interfaces and models. However, it places a burden on students to set up and maintain their own systems.

For students familiar with agentic tools, these software platforms can serve directly as the local RAG client or pipeline\footnote{We have successfully tested this approach using agentic assistants such as Antigravity and OpenCode.}. When instructed to use the structured Markdown repository as a knowledge base, agentic tools automatically utilize rapid text-search commands (such as \texttt{grep}) to scan the directory, isolating specific keywords and contextual patterns relevant to the user's query. Once the search isolates the most pertinent files and exact line numbers, the AI reads those sections to extract the surrounding context. It then synthesizes this raw data into a coherent response, explicitly citing the original file names and line numbers so the user can easily verify the sources.

For students who do not use agentic tools, a simple Python script utilizing scikit-learn and TF-IDF can serve as a highly effective retrieval pipeline. This approach is fast and lightweight, though it lacks the contextual understanding to recognize synonyms or broader abstractions.

\section*{S5: The NuPaD skill set}
\label{sec:skills}

While the NuPaD framework is designed to be simple enough for students to adopt with minimal learning cost, it is complemented by a comprehensive, modular skill set. This architecture rests on a foundational system file, \texttt{NuPaD\_SKILL.md}, which enforces universal pedagogical constraints across all interactions. Operating alongside this core is a suite of nine specialized sub-skills: \texttt{coding-assistant}, \texttt{concept-explainer}, \texttt{data-fetcher}, \texttt{derivation-verifier}, \texttt{homework-assistant}, \texttt{lab-report-builder}, \texttt{literature-reviewer}, \texttt{research-brainstormer}, and \texttt{scientific-writer}. Together, these skills cover most of the activities identified in Tables~\ref{tab:ai_assessments} \& ~\ref{tab:ai_use_cases}.

The entire skill set is designed to load automatically: when a student operates within a modern agentic environment, the system natively detects the directory and loads the foundational rules in the background. As the student engages in a specific task, the environment dynamically invokes the corresponding sub-skill. This seamless integration enforces targeted pedagogical constraints on the fly, ensuring that the AI functions strictly as a regulated educational facilitator without requiring any manual configuration from the student.

\section*{S6: Pedagogical integration in the course}

\subsection{Implementation and pedagogical integration}
\label{sec:evaluation}

The NuPaD framework is designed not to replace lectures but as a complementary educational aid and personalized tutor. Its primary objective is to provide targeted support and help students navigate their specific academic interests, homework assignments, laboratory exercises, long-term projects, and open-ended problems. The framework can also dynamically generate highly engaging, complex problems tailored to individual skill levels, thereby actively promoting deep learning over passive memorization. Specific examples of these generated problems and guided interactions are documented within the textbook repository and the agent skill file. This architecture respects the diverse attitudes that students and educators hold toward artificial intelligence. For those who prefer traditional study methods, the structured Markdown repository serves perfectly well as a standard, high-quality digital textbook. No student is forced to interact with a GenAI. However, for those who seek AI help, the embedded structural metadata ensures that the resulting tutoring experience is rigorous, guided, and pedagogically sound.

\subsubsection*{Personalized tutor and addressing students' background and interests}

The graduate-level subatomic physics course at KTH Royal Institute of Technology presents a pedagogical challenge that is particularly well suited to an AI-mediated solution.
It attracts students from several different subjects including theoretical physics, subatomic and astrophysics, and biomedical physics, with different backgrounds, mathematical fluencies, and disciplinary interests.
In a traditional lecture format, satisfying this diversity simultaneously has been a persistent difficulty.

The NuPaD framework offers a possible solution to this long-standing problem.
Because the interactive agent adapts its questioning depth, choice of analogies, and level of mathematical formalism to the student's self-declared level, and because the structured Markdown textbook explicitly tags content by difficulty level and prerequisite knowledge, the same underlying curriculum can be experienced at different depths by different students.
The tutoring agent does not merely adjust vocabulary; it selects different pedagogical pathways through the same conceptual material, guided by the structural metadata embedded in the textbook.
In this sense, NuPaD offers the possibility of genuinely personalized instruction at scale, refitting the course to each student's background and interests.

\subsubsection*{The peer instruction framework}
In the context of promoting deep learning in the classroom, student-centered activities such as the flipped classroom and the \emph{Peer Instruction} framework~\cite{mazur1997} have been widely proposed as alternatives to passive lecturing. In peer instruction, students are encouraged to think individually and subsequently debate conceptual questions in groups. While this approach is highly successful in introductory courses, our experience indicates that scaling it to advanced master-level curricula presents significant challenges. In advanced courses, student diversity in terms of academic background and specific research interests often becomes the dominant factor. Students frequently express that the format introduces extra stress when grappling with complex, abstract topics under strict time constraints during a live lecture.

GenAI offers a practical mechanism to resolve these bottlenecks in both pre-class preparation and in-class execution. For pre-class preparation, an interactive tutor like NuPaD allows students to engage with challenging conceptual questions at their own pace. Students can iteratively explore their reasoning and build a robust conceptual foundation before the lecture begins. This makes the flipped classroom model viable even for abstract subatomic physics, as the AI systematically bridges the diverse background gaps of individual students.

During in-class discussions, GenAI can further augment the peer instruction dynamic by integrating directly into group debates, acting as an interactive sounding board that proposes alternative physical interpretations. This forces students to critically defend their consensus, thereby elevating the rigor of the peer-to-peer debate. NuPaD can potentially help dynamically generate or adapt questions that specifically target the lingering misconceptions identified during the students' discussions~\cite{weijers2025}.

\subsubsection*{The impact of AI on project-based and take-home assessments}

In physics curricula, take-home assignments, computational projects, and short essays are designed to foster literature exploration, critical evaluation, and the application of knowledge to open-ended problems. The sudden adoption of GenAI drastically alters these activities. Unmediated, AI acts as an answer generator that circumvents independent reasoning and incentivizes superficial learning. While the initial introduction of these tools led to a superficial improvement in the quality of submitted work, it simultaneously compromised evaluative validity. Because it became nearly impossible to distinguish genuine student comprehension from AI-generated text, many universities defensively abandoned take-home projects in favor of heavily invigilated exams~\cite{rudolph2023chatgpt}.

These challenges should not undervalue the pedagogical merits of traditional assessment formats. AI must be structurally constrained to act as an interactive learning partner that supports questioning, reflection, and conceptual exploration. As illustrated in Table~\ref{tab:ai_assessments}, the NuPaD framework transforms these standard activities by replacing the superficial generation of answers with structured, guided inquiry. In addition, educators must rethink the design of these assessments to preserve their pedagogical value. Instead of assigning well-defined algorithmic problems that AI solves effortlessly, assessments must pivot toward ill-defined problems. Students should be tasked with critiquing existing solutions, or intentionally using AI to generate a baseline answer which they must subsequently debug and evaluate.

\begin{table}[hbt!]
\centering
\caption{Contrasting unmanaged AI usage with NuPaD's structured approach across standard physics assessment activities.}
\label{tab:ai_assessments}
\begin{tabular}{>{\raggedright\arraybackslash}p{2.5cm} >{\raggedright\arraybackslash}p{4.5cm} >{\raggedright\arraybackslash}p{5cm}}
\toprule
\textbf{Assessment Activity} & \textbf{Unmanaged Superficial AI Use} & \textbf{NuPaD Structured AI Use} \\
\midrule
Homeworks \& analytical derivations & 
Generates answers from a single prompt, bypassing student struggle. & 
Guides the student step-by-step through inquiry-driven questioning, requiring them to explicitly justify reasoning. \\
\addlinespace
Coding projects & 
Outputs complete code or simulations without requiring student comprehension. & 
Acts as a debugging partner, helping the student trace logic errors while forcing them to write the actual code. \\
\addlinespace
Literature reviews & 
Writes the full essay or report, synthesizing papers the student has never actually read. & 
Serves as an interactive sounding board to brainstorm topics, outline arguments, and critique the student's initial drafts. \\
\bottomrule
\end{tabular}
\end{table}

\subsubsection*{Open-ended problems in subatomic physics education}
\label{sec:wicked-problems}

In advanced subatomic physics, we emphasize open-ended problems in both teaching and evaluations. In examinations, we avoid assigning well-defined, algorithmic tasks, such as standard kinematic derivations or basic matrix diagonalizations, because these are effortlessly solved by AI. Instead, we focus on open-ended problems that require students to synthesize concepts, critique theoretical assumptions, and interpret ambiguous physical data. Students are fully encouraged to answer based on their conceptual understanding, making academic guesses on trivial facts (such as the quark composition of a given particle) if they do not recall them. The NuPaD system is designed to support this shift by acting as a collaborative partner rather than an answer generator.

In defining this pedagogical landscape, we must carefully differentiate between problem classes. Fundamental physics is built upon open-ended problems. Phenomena such as the physical composition of dark matter or the absolute scale of neutrino masses represent open-ended frontiers. They are deeply challenging, yet governed by objective laws of nature: a definitive truth exists, and once the appropriate theoretical frameworks are uncovered, these problems will be solved. Students must therefore view the discipline not as a static repository of finalized textbook answers, but as a continuous progression of effective field theories and empirical models that serve as highly successful approximations of reality.

When we shift our focus from the physical laws to the methodology of teaching itself, however, we immediately encounter a truly ``wicked problem'' introduced by generative AI frameworks. A wicked problem is a sociological challenge characterized by contradictory, shifting requirements that defy permanent solutions. If an academic department aggressively integrates AI to maximize productivity, it risks cultivating a generation of students who struggle with fundamental analytical derivations when the technology is absent. Conversely, a reactionary ban leaves students competitively unequipped for modern research. Because human learning and technical competence are fluidly intertwined, there is no single correct configuration for AI in the classroom; it is a continuous balancing act of structural trade-offs. By explicitly separating this systemic educational challenge from the open-ended pursuit of physical laws, we can train students to maintain their analytical rigor while responsibly navigating the shifting tools of the computational era.

The framework also facilitates a necessary pedagogical shift toward modern computational physics integration.
Rather than requiring students to spend extensive time solving complex equations analytically, advanced courses can use the AI tutor to emphasize the numerical methods and machine learning techniques that dominate contemporary physics research, such as the use of neural networks in data analysis or inverse design.
Historically, integrating these advanced computational techniques into a standard theoretical physics curriculum has been difficult; however, an interactive GenAI tutor can effectively guide students through the coding and implementation phases~\cite{fredly2026how}.

Beyond computational support, the interactive tutor helps clarify obscure theoretical concepts and correct persistent misconceptions in real time.
It can assist in the formulation of interactive simulations and virtual labs, allowing students to visualize abstract subatomic phenomena.
At the graduate level, the AI serves as a collaborative partner to brainstorm novel research directions, synthesize dense academic literature, and engage in dialectic discussions regarding open questions in the field.

\section*{S7: Extended pedagogical rationale}

To reverse the trend of chatbot-induced superficial learning described in Section~\ref{sec:deep-vs-superficial}, we have developed NuPaD, an AI tutoring agent designed explicitly to stimulate deep learning through structured interactions.
In short, the operational principle of NuPaD is to prohibit the GenAI from providing direct 
completed solutions or summary answers on a student's first prompt; instead, it responds like a personalized tutor who guides students through the learning process with step-by-step derivation guidance, interactive problem-solving, and adaptive feedback.

\subsection*{AI-augmented deep and superficial learning and its cautious integration}
\label{sec:deep-vs-superficial}

In higher education, the distinction between deep and superficial learning dictates how students approach and conceptualize knowledge~\cite{biggs2011}. Deep learning\footnote{In this pedagogical context, ``deep learning'' and ``active inquiry'' refer to cognitive comprehension and engaged student participation. These should not be confused with the machine learning / neural network algorithms of deep learning or active learning.} requires students to understand underlying principles and construct robust inferential models, enabling them to evaluate theoretical approximations and transfer conceptual frameworks to novel physics problems. It is characterized by active inquiry (asking ``why?'' and ``how?''), critical evaluation of evidence, and long-term retention, thereby fostering stronger problem-solving skills and creativity.

Conversely, superficial learning prioritizes rapid memorization of isolated facts, formulas, and definitions. It is characterized by passive note-taking, a focus on ``what'' and ``how much,'' and the mechanical repetition of numerical outputs without interrogating the underlying physical boundaries. While such surface-level acquisition is often a necessary initial step in encountering new material, a central objective of advanced physics pedagogy is to transition students beyond rote application toward deep conceptual understanding. This transition is typically encouraged through open-ended problems, research projects, problem-based learning, and interactive peer instruction.

Bloom's Taxonomy classifies educational objectives into a progressive, six-tiered framework designed to push students past passive memorization into deep conceptual mastery~\cite{bloom1956}. A well-designed university physics course must guide students progressively through these cognitive levels: \emph{Remember} (fundamental facts), \emph{Understand} (concepts), \emph{Apply} (principles), \emph{Analyze} (mechanisms), \emph{Evaluate} (evidence), and \emph{Create} (original research). The taxonomy was later revised to focus on dynamic classification to describe cognitive processes, and to introduce a separate knowledge dimension comprising factual, conceptual, procedural, and metacognitive knowledge~\cite{anderson2001}. 
Superficial learning maps to the taxonomy's lowest tiers: remembering facts and understanding basic concepts. True deep learning requires ascending to the higher-order cognitive tiers of analyzing, evaluating, and creating.

The sudden massive usage of conversational GenAI chatbots has introduced a critical challenge into physics education. Fundamentally, GenAI does not inherently promote either deep or superficial learning; rather, it acts as a powerful cognitive amplifier for the student's pre-existing approach. For students already engaged in deep learning, GenAI serves as an accelerator. By automating lower-level tasks, such as generating boilerplate simulation code or retrieving standard equations, AI frees up the student's cognitive capacity to tackle complex analysis and software architecture that would normally be out of reach in a foundational course. 
For superficial learners, GenAI heavily amplifies superficial habits.

Table~\ref{tab:ai_use_cases} outlines common GenAI use-cases in physics education, mapping their potential educational value against their pedagogical risks. The risks associated with unmediated AI usage, such as passive learning, uncritical acceptance of generated code, and deteriorated independent problem-solving, map directly onto the symptoms of superficial learning.  One can state that superficial use of AI amplifies superficial learning, whereas deep and responsible use of AI promotes deep learning.

\begin{table}[htbp]
\centering
\caption{Common generative AI use-cases in physics education. While these tools offer significant educational value through personalized guidance and workflow acceleration, their unmediated use actively promotes superficial learning habits and erodes the critical inquiry required for deep understanding.}
\label{tab:ai_use_cases}
\begin{tabular}{>{\raggedright\arraybackslash}p{3cm} >{\raggedright\arraybackslash}p{4.5cm} >{\raggedright\arraybackslash}p{4.5cm}}
\toprule
\textbf{AI use-case} & \textbf{Educational value} & \textbf{Major risks} \\
\midrule
Conceptual understanding & Explain concepts, compare explanations, answer follow-up questions & Black-box answer machine, passive learning.\\
Problem solving \& tutoring &
Individualized guidance, instant feedback, reinforces conceptual understanding. &
Incorrect reasoning, reduced problem-solving skills. \\
\addlinespace
Programming \& numerical simulation &
Aids debugging, algorithm development, and numerical methods. &
Over-reliance on AI-generated code, shallow programming skills. \\
\addlinespace
Literature review \& reading &
Summarizes papers, explains equations, identifies research gaps. &
Oversimplified interpretations, missing insights, deteriorated reading skills. \\
\addlinespace
Laboratory data analysis &
Data cleaning, plotting, uncertainty analysis, and curve fitting. &
Hidden analysis errors, insufficient validation of workflows. \\
\addlinespace
Report writing \& communication &
Language improvement, \LaTeX{} assistance, streamlined report writing. &
Plagiarism, deteriorated writing skills \& critical thinking. \\
\bottomrule
\end{tabular}
\end{table}

The very features that make AI appealing for education (its ability to provide instant answers, summarize complex topics, and generate coherent text) are the same features that undermine the cognitive processes essential for deep learning. When AI acts as a frictionless answer engine, it circumvents the productive struggle required for conceptual retention. The AI learning paradox can be stated as follows:
\begin{quote}
\textit{The more efficiently an AI system minimizes the cognitive friction required to retrieve, structure, and synthesize knowledge for a learner, the less capable that learner becomes of independently retrieving, structuring, and synthesizing that knowledge in the future.}
\end{quote}

Consider the cognitive dynamics at play when a graduate student encounters a difficult derivation in subatomic physics, such as calculating the transmission probability through the Coulomb barrier via the WKB approximation. The cognitive strain of resolving such an impasse is precisely where genuine neural reorganization and conceptual integration occur~\cite{understanding2024}. When the student engages in productive struggle, formulating hypotheses about valid approximations and debugging flawed reasoning, the resulting understanding is structurally different from knowledge acquired by reading a completed derivation. A student who struggles builds a robust network of inferential connections: an understanding not merely of \emph{what} the answer is, but of \emph{why} certain approximations hold, and \emph{where} they break down.

By bypassing this cognitive resistance entirely, immediate-answer chatbots reduce learning to the passive consumption of synthetic responses. A student who routinely relies on direct answers may reproduce results on short-term assessments, but will remain entirely ill-equipped to identify model boundaries or detect AI fabrications in physical reasoning~\cite{sirnoorkar2024student}. 

\subsubsection*{Responsible use of AI: be creative, critical, practical, and unique}

We must acknowledge that the academic community remains heavily divided on this integration; while some educators enthusiastically embrace AI as a pedagogical revolution, many others remain highly skeptical~\cite{kucukuncular2025teaching,alexandron2026preparing}. A cautious, deliberate approach is therefore essential.

Modern physics curricula must explicitly teach the responsible and productive use of AI. As outlined in Table~\ref{tab:ai_use_cases}, GenAI offers profound opportunities to accelerate the learning process when utilized as an efficient, personalized tutoring system. To maximize these benefits without falling victim to the AI learning paradox, students must be trained to adopt a framework of being creative, critical, practical, and unique.

First, students should use AI to maximize their \emph{creativity}. By offloading the routine friction of learning, such as debugging standard code errors or parsing dense journal syntax, students can redirect their cognitive energy toward high-level theoretical synthesis and experimental design.
Second, students must be relentlessly \emph{critical} of the content the AI produces. Because GenAI models lack genuine physical intuition and are prone to confident fabrications, every generated equation or conceptual explanation must be rigorously interrogated against established physical boundaries. That experience is an invaluable part of the learning and scientific training process.  
Third, students should be ambitious and \emph{practical}. By maximizing the capabilities of AI tools to handle heavy computational tasks, such as generating Monte Carlo scripts, streamlining data analysis workflows, or assisting in the design of complex laboratory setups as noted in Table~\ref{tab:ai_use_cases}, students can tackle problems of far greater complexity than was traditionally possible in a foundational course.
Finally, by using AI as a customized learning and research assistant rather than a uniform answer engine, each student can explore their own specific intellectual curiosities and develop into a \emph{unique}, independent researcher equipped for modern scientific practice.

The relationship between ``mastery'' and ``creativity'' deserves attention. It has always been possible to achieve breakthroughs, even of Nobel Prize level, with a sufficient minimum of core knowledge, provided that knowledge is applied with unique creative insight. The danger is not that AI will replace human creativity; the danger is that AI will erode the very skills that make human creativity possible, or that we fail to use it to enhance them. If the general research and
education community outsources its critical reasoning to GenAI models, the highly literate, competitive ecosystem required to validate and recognize groundbreaking work will disappear, causing scientific progress to stagnate~\cite{birhane2023}.

AI will also change what counts as ``deep learning''.
For now, it presents significant challenges for undergraduate courses, in particular basic courses in physics and engineering science.
For advanced physics courses like subatomic physics, one may find that AI pushes education toward something closer to research practice. Professional physicists already use symbolic algebra systems, numerical software, databases, and increasingly AI assistants. The distinguishing skill is no longer carrying out every calculation by hand, but knowing what question to ask, judging whether the result is physically meaningful, recognizing hidden assumptions, and synthesizing insights across methods. Those are hallmarks of deep learning and may become the primary goals of university physics education in the AI era.

In this context, it may be useful to distinguish between two fundamentally different modes of GenAI-augmented research as well. The first merely utilizes AI as an automated surface tool that accelerates the production of shallow results, substituting computation for comprehension. In a deeper mode, one should use AI to support fundamental discovery through genuinely cognitive activities such as hypothesis generation, equation discovery, and navigating high-dimensional state spaces that are analytically intractable. It may also give the opportunity to strengthen interpretability, enable cross-disciplinary reasoning, and open avenues of inquiry that would otherwise remain inaccessible. Training students to learn and research in this deep mode, rather than defaulting to the first, is precisely the urgent pedagogical challenge in the GenAI era.

\subsubsection*{The case for subatomic physics, an ideal proving ground}
\label{sec:why-subatomic}

Much of the current literature on AI in education focuses on general pedagogy and introductory courses, such as recent studies demonstrating that deploying an off-the-shelf language model as a fallible ``peer'' improves conceptual understanding in basic mechanics~\cite{weijers2025}. GenAI offers uniquely transformative benefits for advanced, graduate-level curricula.

Subatomic physics serves as an ideal proving ground. It is an inherently complex and rapidly evolving discipline characterized by a dense intersection of theoretical formalism, experimental design, and profound open problems. To students, its conceptual landscape often appears as a fragmented collection of disconnected models. In this environment, GenAI acts not as a replacement for the primary lecturer, but as a dedicated personal tutor that helps bridge these conceptual gaps in a manner tailored to the individual student.
Instead of encouraging the mere memorization of textbook facts, a GenAI tutor can train students through rigorous questioning and critical exercise. By continuously challenging students to defend their reasoning and interrogate established theories, the AI cultivates the critical independence essential for future researchers. This pedagogy directly anticipates the evolving nature of scientific work, where routine execution is increasingly automated and human expertise is refocused on problem-finding, orchestration, and high-level theoretical oversight~\cite{bjornson2026}.

The advanced nature of the subject also provides a natural environment for critical evaluation. Rather than passively accepting AI-generated text, graduate-level exercises can require students to actively debug AI outputs. Recent benchmarking efforts demonstrate that even the most advanced frontier models struggle to discover latent physical structures when underlying laws deviate from established science~\cite{wiemann2026}, and similarly fail to reliably solve full-scale, unpublished research challenges~\cite{zhu2025}. This underscores the necessity of training students to empirically verify AI-generated outputs rather than relying on them as established facts.

Table~\ref{tab:bloom_ai} illustrates how the entire taxonomy scales for a master-level subatomic physics course with a cautiously implemented AI-augmented workflow. This course may promote deep learning by guiding students beyond the memorization of nuclear and particle physics facts toward a coherent conceptual understanding of subatomic phenomena. Students are expected to integrate knowledge from quantum mechanics, electromagnetism, relativity, and modern physics to explain, analyze, and interpret physical processes. Rather than simply recalling nuclear properties or particle classifications, students develop the ability to apply physical principles to unfamiliar situations, evaluate experimental evidence, and relate theoretical models to observations.
For instance, rather than simply calculating binding energies (an \emph{Applying} task), students can analyze the deviations between the liquid drop model and raw experimental data to infer the existence of quantum shell closures (an \emph{Analyzing} and \emph{Evaluating} task).
Instead of solving analytical differential equations for a simple linear decay chain, students can build stochastic simulations to model decay networks, observing firsthand how macroscopic deterministic laws emerge from microscopic quantum randomness (a \emph{Creating} task).
By shifting the structural focus toward rigorous model evaluation and open-ended exploration, the educational environment naturally elevates students into the highest cognitive domains.
When carefully supervised, AI facilitates this transition, equipping students with the probabilistic, computational, and critical skills required to contribute to modern scientific research without sacrificing foundational rigor.

\begin{table}[htbp]
\centering
\caption{Bloom's Taxonomy applied to master-level subatomic physics, mapping representative Intended Learning Outcomes and an AI-augmented pedagogical workflow.}
\label{tab:bloom_ai}
\begin{tabular}{l>{\raggedright\arraybackslash}p{5cm}>{\raggedright\arraybackslash}p{5.5cm}}
\toprule
\textbf{Bloom's Level} & \textbf{Representative Learning Outcomes} & \textbf{AI-Augmented Workflow} \\
\midrule
\textbf{Remembering} & Recall Standard Model particle content, decay modes, and basic nuclear properties. & Querying AI to instantly retrieve experimental values and physical constants for direct application. \\
\addlinespace
\textbf{Understanding} & Explain binding energy, shell structure, nucleosynthesis, and detector principles. & Prompting AI to break down dense concepts or explain complex mechanisms via analogies and examples. \\
\addlinespace
\textbf{Applying} & Calculate binding energies; use isotope tables to apply conservation laws to nuclear reactions. & Using AI to rapidly calculate formulas across massive datasets. \\
\addlinespace
\textbf{Analyzing} & Analyze deviations between the liquid drop model and raw experimental data. & Using AI to write scripts that plot model deviations to highlight anomalous trends. \\
\addlinespace
\textbf{Evaluating} & Infer the existence of quantum shell closures; assess validity and limits of physical models. & Using AI as a debating partner to probe theoretical weak points or to draft statistical error-analysis routines. \\
\addlinespace
\textbf{Creating} & Build stochastic simulations to model decay networks; design novel experimental approaches. & Using AI to generate complex Monte Carlo structures while verifying the underlying physics. \\
\bottomrule
\end{tabular}%
\end{table}

\section*{S8: GenAI architecture}

Understanding how GenAI functions is critical for effectively integrating it into physics curricula and avoiding its generation of unphysical claims~\cite{understanding2024} and its amplification of superficial learning.  
There is growing awareness that GenAI introduces obvious inaccuracies, biases, and fabricated facts due to structural limitations in its training data and underlying algorithms. A far more significant threat is the unwarranted overconfidence in the accuracy and objectivity of AI, compounded by low AI literacy among students.
We do not view these technical and cognitive flaws as the primary point of failure for AI in education.
The fundamental pedagogical priority must be to avoid AI's amplification of superficial learning and instead cultivate the student's capacity for deep learning and critical thinking.

The physics research and education community traditionally maintains a cautious approach to technical nomenclature, routinely filtering out peripheral jargon to maintain focus on foundational principles.
The rapid development of artificial intelligence has introduced a dense wave of computational terminology into physics publications, often creating unnecessary confusion.
In this subsection, we establish clear functional definitions for the essential AI concepts utilized in this work.

\subsubsection*{Generative models as function approximators}
In physics, a model is a mathematical framework used to simulate, explain, or predict the behavior of a physical system.
Large language models (LLMs) are analogous, but function as statistical, data-driven systems. At their core, these models are highly sophisticated next-word predictors.
While ``Artificial Intelligence'' is a broad umbrella term spanning everything from simple linear regression to robotics, and ``Generative AI'' specifies tools that create novel content, we focus strictly on LLMs built upon the transformer architecture.
For clarity and to avoid vague generalizations, we will generally drop the qualifiers ``large'' and ``language'' throughout this paper, referring to them simply as the \emph{model}, to bypass debates over size thresholds and avoid the linguistic ambiguity of the term ``language''.

While the fundamental computational unit processed by the architecture is formally known as a \emph{token} (roughly corresponding to a word or sub-word), we intentionally avoid this jargon here.

\subsubsection*{Prompt, agent, and skill: levels of GenAI regulation}

A GenAI model's internal weights are fixed at inference time. Its behavior cannot be altered except through fine-tuning or full retraining. Therefore, the only lever available to a user is what is fed into the model's context window: every instruction, question, uploaded PDF, or reference document constitutes an input prompt. 
When text is fed into a GenAI model, the system tokenizes those words and maps them into a high-dimensional vector space. That input vector alters the internal activation states across billions of parameters, essentially shifting the model's trajectory through its probability distribution. By shaping the prompt, adjusting the phrasing, context, tone, or constraints, one directly manipulates that initial state, guiding the model toward specific regions of its latent space to produce the desired output. The discipline of designing these inputs is popularly called prompt engineering. 
Early GenAI interaction was confined to the chat-interface paradigm, where a single GenAI model generates a response to a prompt. 
There has been substantial progress in shifting from the chat-interface to agentic AI tools.\footnote{Examples of such agentic environments include Claude Code, OpenCode, and Antigravity.} In this paradigm, complex operations are handled by different regulation files (such as skills and agents) operating beneath the model. 

For physics research and education, it is necessary to carefully separate genuinely useful architectures from the commercial incentives of the AI industry. Much of what is marketed as ``agentic architecture'' or ``skill paradigms'' is essentially file organization, determining where and when in the pipeline a piece of text gets inserted. It is effectively I/O scheduling for a GenAI model, dressed in technical terminology. We can distinguish three levels of regulation:

\begin{enumerate}
    \item \textbf{System level, the always-on standing rules.}
    This is the system prompt. It is injected into every turn of the conversation and defines the model's baseline behavior.

    \item \textbf{Workflow level, the modular, on-demand procedure.}
    This is the skill (\texttt{SKILL.md}), a self-contained set of instructions for one specific procedure. A skill is not held in active memory throughout the session; it is retrieved and applied only when the agent recognizes that the current situation matches its trigger condition. This keeps the standing system prompt lean while allowing the agent to draw on an extensible library of specialized behaviors.

    \item \textbf{Task level, the delegated executor.}
    This is the agent (or subagent) in the strict technical sense: a separate, isolated set of execution instructions invoked to carry out a specific task.
    The current generation of agentic development environments extends this into multi-step work with tool-invoked, autonomous agents, using skill files as the modular unit of specialized competence beneath the system prompt. 
    An agent, in this context, is the complete runtime environment: the prompt plus the GenAI plus the available tools (e.g., search engines, compilers, file readers) operating in an active loop.
\end{enumerate}

\subsubsection*{Context engineering and retrieval-augmented generation (RAG)}
\label{sec:rag-intro}

One can state that every layer of interaction with a GenAI model, whether a raw user prompt, an uploaded document, a system message, or a retrieval pipeline, fundamentally serves a single purpose: engineering the final context window and input vector for the static statistical model. From the model's perspective, there is no intrinsic difference between these input sources; all are concatenated into a single word sequence that conditions the next-word probability distribution. While uploaded documents provide high-density local context by placing domain-specific information directly inside the context, RAG acts as a dynamic prompt filter, programmatically searching external storage to inject only the most highly relevant chunks into the context window, augmenting the model's ability to generate an accurate response.
In other words, RAG is composed of retrieval and generation, in addition to pre-processing steps that lead to an embedding model and a vector database.

We conducted a qualitative comparison of existing document indexing and retrieval methodologies across four critical dimensions: mathematical consistency, cross-reference accuracy, token consumption, and local computational overhead, to evaluate how effectively a local GenAI agent can access, contextualize, and reason over textbook content during tutoring interactions.
These three approaches may be the most relevant for university teaching: standard vector Retrieval-Augmented Generation (RAG), hierarchical Wiki RAG platforms, and graph-based indexing (GraphRAG).

In standard vector RAG, the textbook is divided into fixed-size text chunks, each chunk is embedded into a high-dimensional vector space, and at query time the system retrieves the text chunks whose embeddings are most similar to the query embedding.
This approach is computationally minimal and requires only a simple vector database lookup.
This is a good approach for general information retrieval, which is convenient in particular when all lecture notes and slides are provided as a collection of PDF files. 
However, for complex text, fixed-size chunking routinely fractures mathematical equations across chunk boundaries and is incapable of tracking the deep logical relationships between physical variables. RAG has nevertheless been widely deployed in industry and is in particular useful for general-purpose, i.e., non-academic, applications. In practice, the most resource-consuming part is actually the parsing of the documents, which can become challenging when there are hundreds of long documents or when the documents are scanned materials. 

In the Wiki RAG approach, the textbook is organized as a structured hierarchy of files and folders, with each section, subsection, and text block explicitly indexed by its position in the document tree. One can then seamlessly use this Wiki RAG system to augment the GenAI tutoring environment. Retrieval operates by matching the query to the appropriate node in the hierarchy and returning the full section or subsection, preserving the internal coherence of derivations and the boundaries between structural zones.
This approach achieves high cross-reference accuracy, because the index structure explicitly encodes the relationships between sections, and it provides moderate mathematical consistency by retrieving complete derivation blocks rather than arbitrary text fragments.
In practice, the memory and token consumption are moderate, since full sections are retrieved rather than isolated paragraphs, and the computational overhead is low, since retrieval reduces to direct index matching without expensive embedding computations.

Graph-based indexing represents the textbook content as a knowledge graph where physical concepts, equations, variables, and their relationships are encoded with the help of GenAI.
It retrieves not only the directly relevant content but also the network of related concepts. It can be quite powerful for cross-indexing over a large number of documents.
However, graph construction depends on a GenAI-driven extraction pass across the entire textbook, which is inherently lossy: relationships may be missed, misattributed, or fabricated, and mathematical structure can be fragmented when equations are decomposed into discrete structural nodes rather than retrieved as intact derivations. The computational cost is also significant: constructing the graph requires heavy pre-computation of entity extraction and relationship identification, and querying requires traversal of structural subgraphs and aggregation of entity summaries, resulting in both high token consumption and substantial local computational overhead.

Table~\ref{tab:indexing} summarizes the comparative evaluation across the four critical dimensions.

\begin{table}[t]
\caption{\label{tab:indexing}Comparison of document indexing and retrieval methodologies for advanced physics curricula. We term Wiki RAG as a hierarchical framework that transforms documents into a tree index RAG and then synthesizes a Wikipedia-like interlinked indexing page with or without help from GenAI.}
\begin{tabular}{@{}>{\raggedright\arraybackslash}p{2.5cm}>{\raggedright\arraybackslash}p{3cm}>{\raggedright\arraybackslash}p{3cm}>{\raggedright\arraybackslash}p{3cm}@{}}
\toprule
Metric & Vector RAG & Wiki RAG & GraphRAG \\
\midrule
Math consistency & Poor; equations fractured across chunks & Excellent; equations retrieved intact within complete sections & Variable;  loss of derivation context despite explicit variable linkage \\
Cross-reference accuracy & Very low; fails to link distant chapters & High; explicit index networks & Potentially strong but GenAI extraction-dependent \\
Local compute overhead & Minimal; vector lookup & Low; direct index matching & Significant; heavy pre-computation\\
\bottomrule
\end{tabular}
\end{table}

The advantage of RAG is that the effective knowledge available to the GenAI is not limited by the model context window. One recent framework~\cite{asai2026}, for instance, was developed as a retrieval-augmented approach for scientific literature synthesis.
Their work does not introduce a new language model architecture; rather, it improves the retrieval and evidence-selection pipeline for scientific documents.
The system performs multi-stage retrieval, document ranking, passage selection, and evidence aggregation before generating a literature synthesis using a GenAI.
The main limitation is that the quality of the final response depends on the retrieval process: information that is not retrieved cannot contribute to the generated answer.

\subsubsection*{Markdown file format}
Markdown can be considered the native format for GenAI. From an architectural perspective, it serves as a simplified, lightweight alternative to \LaTeX{} that acts as an exceptionally token-efficient, clean database for several interconnected reasons:
Markdown relies on minimal inline syntax to convey a rich document hierarchy with minimal token overhead. Because modern GenAI are trained extensively on web content and code repositories, they understand these lightweight structural primitives perfectly. Consequently, Markdown provides an optimal balance between structural precision and token efficiency, allowing GenAI to parse document layout and domain knowledge effortlessly.
Unlike \LaTeX{} source files, which contain extensive compilation information that consumes large portions of the GenAI's context window with non-essential noise, Markdown contains virtually zero visual layout overhead. This ensures that a far greater proportion of the token budget is devoted to actual physics content.
Markdown's hierarchical header system  provides a clean tree structure that RAG systems can use to divide the textbook into logically coherent chunks without breaking equations, derivation sequences, or code blocks.

For the specific use case of local GenAI tutoring in subatomic physics, we find that a structured  Markdown database combined with hierarchical indexing in the Wiki RAG format provides the most effective balance: it achieves high cross-reference accuracy through the explicit index structure and custom XML tags described in Section~\ref{sec:markdown}, maintains strict mathematical consistency by retrieving complete logical blocks rather than arbitrary text fragments, and operates with low token overhead and minimal computational cost.

\subsection*{Model inference and local deployments}

In AI terminology, inference refers to the computational act of executing a trained model to generate a response. During this process, the computer loads billions of model parameters into memory and performs massive matrix multiplications. While commercial models rely on massive data centers for this task, NuPaD uses open-source inference engines\footnote{Such as \texttt{llama.cpp}, an open-source C/C++ inference engine.} that run models natively on standard consumer hardware, such as a laptop or a local university workstation. In this local deployment, the model weights are stored as static data files and loaded directly into the inference engine.
\subsubsection*{Cloud and local models}

Cloud-based models are massive systems hosted on commercial server infrastructure.
They offer powerful reasoning capabilities and user-friendly interfaces with minimal setup.
Their reliance on constant internet connectivity and external data transmission raises significant student data privacy and security compliance issues.

A locally deployed GenAI is executed entirely on a user's physical machine, such as a laptop or a local university workstation.
Local models are typically quantized to fit within the stringent memory constraints of consumer hardware.
While this may sacrifice some raw reasoning capacity compared to cloud counterparts, local execution ensures complete data privacy, offline availability, and zero subscription or API costs.
Because these models are largely open-source, they permit direct programmatic control.

While local models are broadly categorized into dense architectures and Mixture-of-Experts (MoE) variants, MoE variants can substantially reduce per-token inference cost relative to similarly sized dense models, since only a fraction of the total parameters is activated for any given token. There are also non-Transformer-based AI models such as Mamba, which is a state space model that has shown promising results in terms of inference efficiency, but their performance is not yet satisfactory for high-level physics education.

In the context of educational research and the development of NuPaD's algorithms, it is essential to utilize open-source local models rather than opaque commercial black boxes. The adoption of open-source AI is critical not only for safeguarding ethics, data privacy, and institutional values, but also for ensuring equitable availability and the freedom to adapt educational tools to diverse pedagogical contexts~\cite{unesco2023guidance}.
While the AI industry continuously pursues extremely large models to top generalized benchmarks, the specialized pedagogical demands of physics education demonstrate that smaller, highly optimized local models are not only sufficiently capable, but often superior in flexibility, privacy, and cost-effectiveness.

\subsubsection*{Interfaces and tooling}
GenAI first gained significant adoption due to its text generation capabilities in chatbot-style web interfaces or phone applications. These interfaces are highly intuitive and allow users to interact with GenAI using natural language prompts. Local deployments can utilize chat-based interfaces through  tools such as Open WebUI.
The GenAI landscape is rapidly evolving from simple text interfaces into integrated computational workflows. 
Behind the scenes, the user interacts with a GenAI through a sequence of complex tools organized by the interface, which may be hidden from the user to provide a smoother experience, but they perform substantial pre-engineering before the input question is finally sent to the GenAI.

\putbib[references]
\end{bibunit}

\end{document}